\documentclass[12pt]{article}

\usepackage[a4paper,total={6.5in,9.5in}]{geometry}
\usepackage[utf8]{inputenc}
\usepackage[colorlinks=True,
    linkcolor=Cerulean,
    filecolor=Cerulean,
    urlcolor=Cerulean,
    citecolor=Cerulean]{hyperref}

\usepackage{mathptmx}

\usepackage{titlesec}
\titleformat{\section}{\large\bfseries}{\thesection.}{0.5em}{}  %
\titleformat{\subsection}{\large\itshape}{\thesubsection.}{0.5em}{}
\titleformat{\subsubsection}{\itshape}{\thesubsubsection.}{0.5em}{}
\titleformat{\paragraph}{\itshape}{\theparagraph}{0.5em}{}

\usepackage{indentfirst}
\usepackage{setspace}
\usepackage{textcomp}

\usepackage{siunitx}
\usepackage{bm}
\usepackage{amsmath}
\usepackage{amsfonts}
\usepackage{amssymb}
\newcommand\independent{\protect\mathpalette{\protect\independenT}{\perp}}
\def\independenT#1#2{\mathrel{\rlap{$#1#2$}\mkern2mu{#1#2}}}

\usepackage[dvipsnames,table]{xcolor}
\usepackage{graphicx}
\graphicspath{{./images/}}
\definecolor{battleshipgrey}{rgb}{0.52, 0.52, 0.51}
\usepackage{rotating}

\usepackage{tabularx}
\usepackage{multirow}
\usepackage{array}
\usepackage{booktabs}
\usepackage{caption}
\usepackage{subcaption}
\usepackage[bottom,multiple,symbol]{footmisc} 
\usepackage{bookmark}

\usepackage[american]{babel}
\usepackage{csquotes}

\usepackage{xpatch}

\xpretobibmacro{date+extradate}{\unspace\addcomma\addspace}{}{}

\makeatletter
\renewcommand\p@figure{Fig. }
\renewcommand\p@table{Table \arabic{table}\expandafter\@gobble}
\renewcommand\p@section{Section \arabic{section}\expandafter\@gobble}
\makeatother

\usepackage{pgfplots}
\pgfplotsset{compat=newest}

\usepackage{tikz}
\usetikzlibrary{arrows,decorations.pathmorphing,backgrounds,positioning,fit,petri}
\usetikzlibrary{bayesnet}
\usetikzlibrary{mindmap,trees}
\usetikzlibrary{positioning}

\begin{document}

\title{\vspace{-2ex}\fontsize{20}{22}\selectfont Uncovering life-course patterns with causal discovery and survival analysis \vspace{-2ex}}

\author{}
\date{}
\maketitle

\begin{center}
    \fontsize{12}{14}\selectfont \textbf{Bojan Kostic}\textsuperscript{a,}\footnote{Corresponding author. E-mail address: \href{mailto:boko@dtu.dk}{boko@dtu.dk} (B. Kostic).},
    \textbf{Romain Crastes dit Sourd}\textsuperscript{b},
    \textbf{Stephane Hess}\textsuperscript{c},
    \textbf{Joachim Scheiner}\textsuperscript{d},
    \textbf{Christian Holz-Rau}\textsuperscript{d},
    \textbf{Francisco C. Pereira}\textsuperscript{a}
\end{center}
\vspace{3ex}

{\fontsize{11}{13}\selectfont \textsuperscript{a}Machine Learning for Smart Mobility (MLSM) Group, Transport Division, Department of Technology, Management and Economics, Technical University of Denmark, Lyngby, Denmark

\textsuperscript{b}Centre for Decision Research, Leeds University Business School, University of Leeds, Leeds, UK

\textsuperscript{c}Choice Modelling Centre, Institute for Transport Studies, University of Leeds, Leeds, UK

\textsuperscript{d}Department of Transport Planning, Faculty of Spatial Planning, TU Dortmund, Dortmund, Germany}
\hfill \break
\\

\textbf{Abstract}.
We provide a novel approach and an exploratory study for modelling life event choices and occurrence from a probabilistic perspective through causal discovery and survival analysis. Our approach is formulated as a bi-level problem. In the upper level, we build the life events graph, using causal discovery tools. In the lower level, for the pairs of life events, time-to-event modelling through survival analysis is applied to model time-dependent transition probabilities. Several life events were analysed, such as getting married, buying a new car, child birth, home relocation and divorce, together with the socio-demographic attributes for survival modelling, some of which are age, nationality, number of children, number of cars and home ownership. The data originates from a survey conducted in Dortmund, Germany, with the questionnaire containing a series of retrospective questions about residential and employment biography, travel behaviour and holiday trips, as well as socio-economic characteristic. Although survival analysis has been used in the past to analyse life-course data, this is the first time that a bi-level model has been formulated. The inclusion of a causal discovery algorithm in the upper-level allows us to first identify causal relationships between life-course events and then understand the factors that might influence transition rates between events. This is very different from more classic choice models where causal relationships are subject to expert interpretations based on model results. \\


\setlength{\parindent}{1cm}
\section{Introduction and Background}
\label{sec:introduction}


Inspired by a few seminal studies \cite{hagerstrand1970people, chapin1974human} highlighting how different life-course decisions are interrelated and should be analysed jointly, several studies \cite{muggenburg2015mobility} have analysed travel behaviour as a long-term life-course dimension, similar to employment spells or residential location choices. Retrospective surveys such as the one proposed by Beige and Axhausen \cite{beige12} seek to uncover how life events, such as the birth of a child or marriage, are connected to one another as well as to long term mobility decisions such as commuting mode or relocation. Retrospective surveys are generally based on a life-course calendar, where the respondents are asked a series of questions for each year in retrospect (which is a grid analogous to Figure A1). Life-course calendars are commonly used to collect such information where panel data is unavailable or too costly. Life-course calendars have become popular because they have two main advantages over traditional panel surveys: the visual and mental relation of different kinds of events improves the quality of the data by providing reference points and preventing time inconsistencies. Moreover, the visual aid of the calendar tool eases the task of listing a potentially high number of different short events, a task that would be more difficult in traditional surveys \cite{freedman1988life}. However, while scholars agree on the fact that life-course calendars provide a helpful tool to facilitate the recall of life events, many acknowledge the issue of recall of all relevant events and the effort required to do so. In particular, Manzoni et al. \cite{manzoni20102} and Manzoni \cite{manzoni2012and} find that the recall depends not only on respondents' socio-demographic characteristics but also on the type of events that are recalled. For example, people tend to omit or reduce the duration of spells of unemployment. In addition, it has been shown by Schoenduwe et al. \cite{schoenduwe2015analysing} that participants find it easiest to remember ``emotionally meaningful, momentous, unique or unexpected events''. It is precisely these kinds of events that we consider in our analysis.

The data for our research originates from a retrospective survey that was carried out between 2007 and 2012 at the Department of Transport Planning of the Faculty of Spatial Planning at TU Dortmund. The questionnaire contains a series of retrospective questions about residential and employment biography, travel behaviour and holiday trips, as well as socio-economic characteristics.

In our work, we try to investigate if causal relations between life events can be identified in our dataset, such that these can then be used for choice modelling and provide valuable insights into understanding people's behaviour. We do this through causal discovery, a tool that deals with the process of automatically finding, from available data, the causal relations between the available (random) variables. This typically follows a graphical representation, where nodes are the variables, and arcs are the causal relationships between them. A causal discovery algorithm thus has as main objective to find directed graphical causal models (DGCMs) that reflect the data. There is also research on acyclic graphs and combinations of both, which we will not focus on here. For a broad review, we recommend Glymour et al. \cite{causal_discovery_review}.

The ideal way to discover causal relations is through interventions or randomized experiments. However, in many cases, such as ours, this is in fact impossible. Having only pure observational data, causal discovery tools search for systematic statistical properties, such as conditional independence, to establish (possible) causal relations. This implies also that we are limited to what the data \emph{can tell} us about causality. If the true causality structure is not observable from the data, such methods will at most find correlations disguised as causation, as happens in many other non-causal methods, such as regression. In such cases, a possible solution relies on domain knowledge, such as postulating the existence of common causes (e.g. a latent variable), or the addition of constraints.

The concept of causal discovery thus consists of algorithms to discover causal relations by analyzing statistical properties of purely observational data. In this paper, we intend to apply such methods to automatically uncover causality through temporal sequence data. Time or sequence are hardly a true cause in decision making (e.g. one will not have children or buy a car just because X time has passed, but due to  some deeper cause). Therefore, it is certainly a bit too bold to claim that there is inherent causality in all the relationships that we will find. Instead, we see this process as uncovering the most likely graph of sequential patterns throughout the life of individuals.

Given a resulting causal graph, together with a dataset of temporal sequence records, we can now estimate transition probabilities for each event pair from the graph. We model the transition probabilities through survival analysis, which is typically used for time-to-event (TTE) modelling (or similarly for time-to-failure (TTF) or durations) \cite{survival_analysis_2}. Classical survival analysis is widely applied in medical studies to describe the survival of patients or cancer recurrence \cite{10.1371/journal.pbio.0020108, aalen-history, Cheng181ra50, Royston2013, 10.1001/jama.2016.3775}. Beside medical applications, it is now applied far beyond this field, such as predictive maintenance, customer churn, credit risk, etc.

Through survival analysis, given that an initial event occurred, we can obtain probabilistic temporal estimates of another event occurrence, with the survival values defined as $ S(t) = \mathrm{P}(T > t) $. It represents the probability of the event happening after time $ t $ from the initial event (i.e. the probability of surviving past time $ t $), given that the event did not happen before time $ t $.

In our case, survival probabilities will give us estimates of how likely it is that a certain life event will occur in each coming year after another specific life event. This leads to a better understanding of people's life-course calendars and how these depend on their socio-demographics. Examples include how the number of cars affects a decision of moving after getting married, the effect of the number of children on buying a car after moving, the difference between homeowners and non-homeowners when it comes to having a child after wedding, to name a few.

\section{Methods}
\label{sec:methods}

Our approach is formulated as a bi-level problem: in the upper level, we build the life events graph, with causal discovery tools, where each node represents a life event; in the lower level, for selected pairs of life events, time-to-event modelling through survival analysis was applied to model the time-dependent probabilities of event occurrence.

\subsection{Causal Discovery}
Two causal discovery algorithms used in our research are Parents and Children (PC) and Greedy Equivalence Search (GES). In a resulting life events graph, we should have a set of key events as nodes, while the arcs correspond to temporal ``causality'' between pairs of events, where $ A \rightarrow B $ implies that $ A $ is precedent to $ B $. As discussed earlier, this is not assumed as true causality; instead, we see it as representing a high likelihood of observing a certain temporal order in such events.

The concept of conditional independence is key here:

\[ X \independent Y | S \,\, \iff \,\, P(X, Y|S) = P(X|S)P(Y|S) \,\,\textit{for all dataset} \]

The idea is quite intuitive: if two variables $ X $ and $ Y $ are independent when controlling for the (set of) variable(s) $ S $, then $ X $ and $ Y $ are conditionally independent given $ S $. This means that $ S $ is likely to be a common cause for $ X $ and $ Y $. Of course, just as testing for independence in general, testing for conditional independence is a hard problem \cite{shah2018hardness}, but approximate techniques exist to determine it from observational data \cite{dawid1979conditional}.

Algorithms such as Parents and Children (PC) or its derivatives such as the Fast Causal Inference (FCI) \cite{spirtes2000causation} are based on this principle to generate a causal graph. They essentially start from a fully connected undirected graph, and gradually ``cut'' links between conditionally independent nodes.

Another approach starts from the opposite direction, where we start from an empty graph, and gradually add nodes and arcs. The most well-known algorithm is the Greedy Equivalence Search (GES) \cite{chickering2002optimal}. At each step in GES, a decision is made as to whether adding a directed edge to the graph will increase fit measured by some fitting score (e.g. Bayesian information criterion (BIC), z-score). When the score can no longer be improved, the process reverses, by checking whether removing edges will improve the score. It is difficult to compare this with PC in terms of the end quality of the discovered graph, but it is known that, as the sample size increases, the two algorithms converge to the same Markov Equivalence Class \cite{causal_discovery_review}. The vast majority of causal discovery algorithms are variations or combinations of the two approaches just mentioned.

In our dataset, we have four events, namely \emph{child birth}, \emph{wedding}, \emph{new car} and \emph{moving}, and two state variables, namely \emph{married} (True or False) and \emph{children} (numeric). While the final life events graph shall have only events (the state variables will be useful later, but not as nodes in the graph), PC and GES allow for inclusion of these variables, therefore we applied both PC and GES algorithms \cite{kalainathan2019causal}.

\subsection{Survival Analysis}
In survival analysis, survival function $ S(t) $ can be defined from the cumulative distribution function $ F(t) $, as $ S(t) = 1 - F(t) $, which in terms of probabilities can be formulated as $ S(t) = \mathrm{P}(T > t) $. The survival function can be interpreted in terms of the gradient (slope) of a line, with steeper lines meaning an event is more likely to happen in the corresponding interval. In contrast, a flat line means there is a zero probability of an event happening in that interval.

Most commonly used statistical methods for survival analysis are the Kaplan-Meier method \cite{kaplan-meier} and the Nelson-Aalen method \cite{nelson, aalen}, which are univariate counting-based methods for general insights. To estimate the effect of covariates, a survival regression model can be formalized through the Cox proportional hazards model (CPH) \cite{coxph}. It assumes that all individuals share the same baseline hazard function, and the effect of covariates acts as its scaler. It provides interpretability but lacks the accuracy with larger datasets.

In survival analysis, time-to-event (TTE) is expressed through two events: the \textit{birth} event and the \textit{death} event (hence the term ``survival''). Time between the two events is survival time, denoted by $ T $, which is treated as a non-negative random variable. Given the \textit{birth} event, it represents the probability distribution of the \textit{death} event happening in time.

Observations of true survival times may not always be available, thus creating distinction between uncensored (true survival times available) and censored observations (true survival times unknown, but only the censorship time). This information is expressed through an event indicator $ E $, with binary outcome. The censorship time is assumed to be non-informative, and $ T $ and $ E $ are independent.

Two main outputs of survival analysis are the survival function and the (cumulative) hazard function. The survival function $ S(t) $ can be formulated starting from the cumulative distribution function (CDF), denoted by $ F $, which can be defined as the probability (\textrm{P}) of an event occurring up to time $ t $:
\begin{equation}
    \label{eq:cdf}
    F(t) = \mathrm{P}(T \le t) = \int_{0}^{t} f(x) dx
\end{equation}
with properties of being a non-decreasing function of time $ t $, having bounds on the probability as $ 0 \le F(t) \le 1 $, and its relation with the survival function as $ F(t) = 1 - S(t) $. Given this, we can define the survival function as the probability of surviving past time $ t $, i.e. the probability of an event happening after time $ t $ given that the event has not happened before or at time $ t $ as:
\begin{equation}
    \label{eq:surv}
    S(t) = \mathrm{P}(T > t) = \int_{t}^{\infty} f(x) dx
\end{equation}
with properties of being a non-increasing function of $ t $, and having bounds on the survival probability as $ 0 \le S(t) \le 1 $ and $ S(\infty) = 0 $, given its relation with the CDF as $ S(t) = 1 - F(t) $. The probability density function (PDF) of $ T $, $ f(t) $, can be formalized as: 
\begin{equation}
    \label{eq:surv-pdf}
    f(t) = \lim_{\Delta t \rightarrow 0} \frac{\mathrm{P}(t < T \le t + \Delta t)}{\Delta t}
\end{equation}

The hazard function $ h(t) $ defines the instantaneous hazard rates of the event happening at time $ t $, given that the event has not happened before time $ t $. It can be defined as a limit when a small interval $ \Delta t $ approaches 0, with a probability of the event happening between time $ t $ and $ t + \Delta t $, as in \eqref{eq:hazard-def}. In addition, there is a direct correlation between the hazard function and the survival function.
\begin{equation}
    \label{eq:hazard-def}
    h(t) = 
    \lim_{\Delta t \rightarrow 0} \frac{\mathrm{P}(t < T \le t + \Delta t \left\vert\right. T > t)}{\Delta t} =
    \frac{f(t)}{S(t)} =
    -\frac{d}{dt} \log{(S(t))}
\end{equation}
The cumulative hazard function $ H(t) $ is also used, and is especially important for model estimation \eqref{eq:cum-hazard}. The survival function can therefore be derived by solving the differential equation above, which gives the relation with the hazard and cumulative hazard functions and is expressed in \eqref{eq:surv-hazard}.
\begin{equation}
    \label{eq:cum-hazard}
    H(t) = \int_{0}^{t} h(x) dx = - \log{(S(t))}
\end{equation}
\begin{equation}
    \label{eq:surv-hazard}
    S(t) = \exp{\left( -\int_{0}^{t} h(z) dz \right)} = \exp{(-H(t))}
\end{equation}

Survival analysis is typically conducted through univariate survival models or through survival regression. The former takes into account only survival times, while the latter also includes covariates. The most widely used regression model is the Cox proportional hazards (CPH) model \cite{coxph}. In addition, there have been recent developments in applying machine learning models in survival analysis to improve prediction accuracy, but compromising on model interpretability. Machine learning algorithms, such as support vector machines \cite{surv-svm} and random forests \cite{surv-rf} are common examples of such techniques. Beside better accuracy, they also demonstrated a superior efficiency over existing training algorithms, especially for large datasets.

To exploit non-linear relationship among the covariates, neural networks (NN) \cite{francisco-book-ch2} are used. The first time a simple neural network was proposed was by Faraggi and Simon \cite{faraggi}. The approach is similar to the CPH, but optimizes a slightly different partial likelihood function. More recent developments are the application of deep neural networks (DNNs), with models like DeepSurv \cite{deepsurv}, DeepHit \cite{deephit} and Nnet-survival \cite{nnet-survival}. DeepSurv is a proportional hazards deep neural network. It is an extension of the method proposed by \cite{faraggi}, used to connect more than one layer in the network structure. While CPH and DeepSurv rely on the strong proportional hazards assumption, other models relax this condition. DeepHit uses a DNN to learn the survival distribution directly. It imposes no assumptions regarding the underlying processes and allows for the effect of the covariates to change over time. Nnet-survival, on the other hand, is a discrete time survival model. The hazard rate, which for the previously discussed methods is calculated as the instantaneous risk, here is defined for a specific time interval.

Because of the relatively small dataset for deep learning applications and to focus on the interpretability, in this paper we used the univariate models and the CPH regression model.

\subsubsection{Univariate survival models}
\label{subsubsec:method-univariate-surv-models}

The estimation of the survival and cumulative hazard functions can be performed using non-parametric univariate models. Survival and cumulative hazard functions are functions of survival times with event indicator, i.e. $ S(t) = f(\mathrm{\mathbf{T}}, \mathrm{\mathbf{E}}) $ and $ H(t) = g(\mathrm{\mathbf{T}}, \mathrm{\mathbf{E}}) $. Given the ordered survival times, for each instant of interest on the timeline, the number of individuals at risk and the number of death events occurring is determined. The survival function is usually estimated using the Kaplan-Meier method \cite{kaplan-meier}, a product-limit estimator. It takes as input survival times and estimates the survival function as:
\begin{equation}
    \label{eq:km}
    \hat{S}(t) = \prod_{i: t_i \le t} \left( 1 - \frac{d_i}{n_i} \right)
\end{equation}
where $ d_i $ is the number of death events at time $ t $ and $ n_i $ is the number of individuals at risk of death just prior and at time $ t $. Similarly, the cumulative hazard function can be estimated using the Nelson-Aalen method \cite{nelson, aalen} as:
\begin{equation}
    \label{eq:na}
    \hat{H}(t) = \sum_{i: t_i \leq t} \frac{d_i}{n_i}
\end{equation}
Hazard and cumulative hazard values cannot be directly interpreted, but are used for relative comparison among themselves in discovering which timeline instants are more likely to encounter a death event than others.

Univariate non-parametric survival models are simple and can provide flexible survival and hazard curves. They provide population-level insight into survival behaviour, but not at the individual level.

\subsubsection{Survival regression}
\label{subsubsec:method-survival-regression}

To account for the effect of covariates on the output functions, the Cox proportional hazards model is typically used for survival regression. It is a semi-parametric regression model that estimates the risk, i.e. the log of the hazard rate, of an event occurring as a linear combination of the covariates as $ \bm{\beta}^{\mathsf{T}} \mathrm{\mathbf{x}} $, with $ \bm{\beta} $ being a vector of coefficients and $ \mathrm{\mathbf{x}} $ denoting a vector of time-constant covariates. It maximizes the Cox's partial likelihood function (of the coefficients) \cite{coxph}:
\begin{equation}
    \label{eq:coxph_likelihood}
    L(\bm{\beta}) = \prod_{i: E_i = 1} \frac{\exp \left( \bm{\beta}^{\mathsf{T}} \mathrm{\mathbf{x}}_i \right)}{\sum_{j: T_j \ge T_i} \exp \left( \bm{\beta}^{\mathsf{T}} \mathrm{\mathbf{x}}_j \right)}
\end{equation}
where only uncensored data ($ i: E_i = 1 $) are used (hence the term ``partial''). The corresponding negative log-likelihood function for minimization is given by \eqref{eq:coxph_loglikelihood}. As implemented in \cite{lifelines}, the optimization problem is solved using maximum partial likelihood estimation (MPLE) and Newton-Raphson iterative search method \cite{suli_mayers_2003}. Ties is survival times are handled using the Efron method \cite{efron}. The L2 regularization is used with a scale value of 0.1.
\begin{equation}
    \label{eq:coxph_loglikelihood}
    l(\bm{\beta}) = - \log L(\bm{\beta}) = - \sum_{i: E_i = 1} \left( \bm{\beta}^{\mathsf{T}} \mathrm{\mathbf{x}}_i - \log \sum_{j: T_j \ge T_i} \exp \left( \bm{\beta}^{\mathsf{T}} \mathrm{\mathbf{x}}_j \right) \right)
\end{equation}

The hazard function is computed for each individual as a product of the population-level time-dependent baseline hazard $ h_0(t) $ and the individual-level time-constant partial hazard expressed through the exponent of a linear combination of the covariates. It is formulated as:
\begin{equation}
    \label{eq:coxph_hazard}
    h(t \vert \mathrm{\mathbf{x}}) = h_0(t) \exp \left( \bm{\beta}^{\mathsf{T}} \mathrm{\mathbf{x}} \right)
\end{equation}
All individuals share the same baseline hazard, and the effect of covariates acts as a scaler through the partial hazard. The baseline hazard function is estimated using the Breslow method \cite{breslow}. The exponential form ensures that the hazard rates are always positive. The ratio of the hazard functions of two individuals is always constant and it is given by the ratio of their partial hazards (hence the term ``proportional''). Given the hazard function, cumulative hazard function and survival function can be computed using \eqref{eq:cum-hazard} and \eqref{eq:surv-hazard}, respectively.

The goodness-of-fit measure is the concordance index (C-index) \cite{regr-model-book}, a commonly used metric to validate the predictive capability of survival models (analogous to the $ R^2 $ score in regression). It is a rank correlation score, comparing the order of observed survival times from the data and order of estimated survival times (or negative hazard rates) from the model. It is formulated as:
\begin{equation}
    \label{eq:c_index}
    C = \frac{1}{\epsilon} \sum_{i: E_i = 1} \sum_{j: T_j \ge T_i} \mathbf{1}_{f(\mathrm{\mathbf{x}}_i) < f(\mathrm{\mathbf{x}}_j)}
\end{equation}
where $ \epsilon $ is the number of samples used in calculation and $ \mathbf{1} $ is the indicator function with binary outcome based on the given condition. The C-index is equivalent to the area under a ROC (Receiver Operating Characteristic) curve, and ranges from \numrange{0}{1}, with values less than 0.5 representing a poor model, 0.5 is equivalent to a random guess, \numrange{0.65}{0.75} meaning a good model, and greater than 0.75 indicating a strong model, with 1 being the perfect concordance and 0 being the perfect anti-concordance.

\section{Data elaboration and model fitting}
\label{sec:data}

The data used in model estimation contains both uncensored and censored observations. Uncensored observations for an individual are survival times available directly as the number of years between two specific events, with the event observed indicator set to 1. Censored observations are created when only the birth event happened, with survival times set to 20 years and the event observed indicator set to 0. 

To fit the models and validate their performance, the dataset, which comprises 1,487 individuals, was split into the training set, containing \SI{60}{\percent} of the samples, and the test set, as the hold-out set containing the remaining \SI{40}{\percent} of the samples. Stratified splitting was used to ensure a balanced presence of covariates' values in both datasets.

Data elaboration and modelling was performed in Python \cite{python}, using packages for numerical and scientific computing, data manipulation and machine learning \cite{pandas, scikit-learn, numpy, scipy}. The visualizations were created using Python plotting libraries \cite{matplotlib}. Survival modelling was done in the lifelines library \cite{lifelines}.

An example of elaborated time-dependent data for one person from the dataset used for survival analysis is shown in Table A2.

\subsection{Covariates description}

The covariates are divided into five groups: general attributes, independent state attributes, dependent state attributes, transition attributes and indicator transition attributes. Their description, type and feasible values are shown in Table A3.

\subsection{Model fitting}

In survival regression, survival and cumulative hazard functions are functions of survival times with event indicator and covariates with corresponding coefficients, $ S(t) = f(\mathrm{\mathbf{T}}, \mathrm{\mathbf{E}}, \mathrm{\mathbf{x}}, \bm{\beta}) $ and $ H(t) = g(\mathrm{\mathbf{T}}, \mathrm{\mathbf{E}}, \mathrm{\mathbf{x}}, \bm{\beta}) $. Transition attributes are used as states for the analysis, whereas the remaining ones are used for conditioning. Table A4 shows results of fitting a CPH model for each pair of life events from the causality graph. The table also contains information on the number of existing samples for each pair, out of which, as mentioned above, 60 \% were used for fitting and 40 \% as hold-out data for validation.

\section{Results and Discussion}
\label{sec:results}

\subsection{Life Events Graph}

\begin{figure}[h!]
    \centering
    \includegraphics[width=0.65\linewidth]{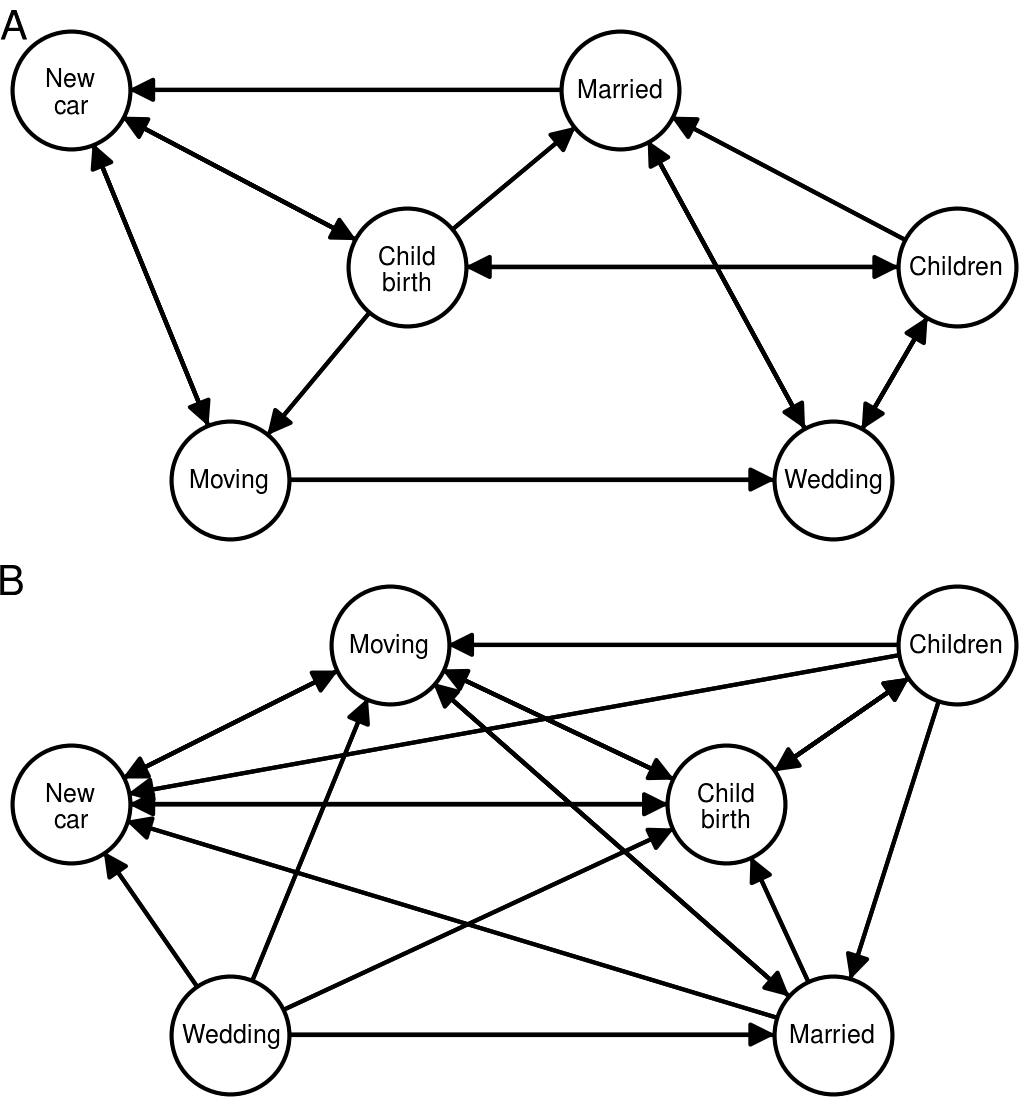}
    \caption{Life events graphs created using PC and GES algorithms. (\textit{A}) PC. (\textit{B}) GES.}
    \label{fig:pc-ges}
\end{figure}

To build our graphs, we ran Parents and Children (PC) and Greedy Equivalence Search (GES) algorithms with prepared data. Not surprisingly, we verified that their final graphs were not mutually consistent in several components. \ref{fig:pc-ges} shows the results. We can also see that many event pairs have bidirectional arrows. This is neither surprising nor necessarily problematic, it only means that temporal precedence can flow in either direction, and there may be cyclic stages in life (child birth $ \rightarrow $ moving $ \rightarrow $ new car...). 

\begin{figure}[h!]
    \centering
    \includegraphics[width=0.65\linewidth]{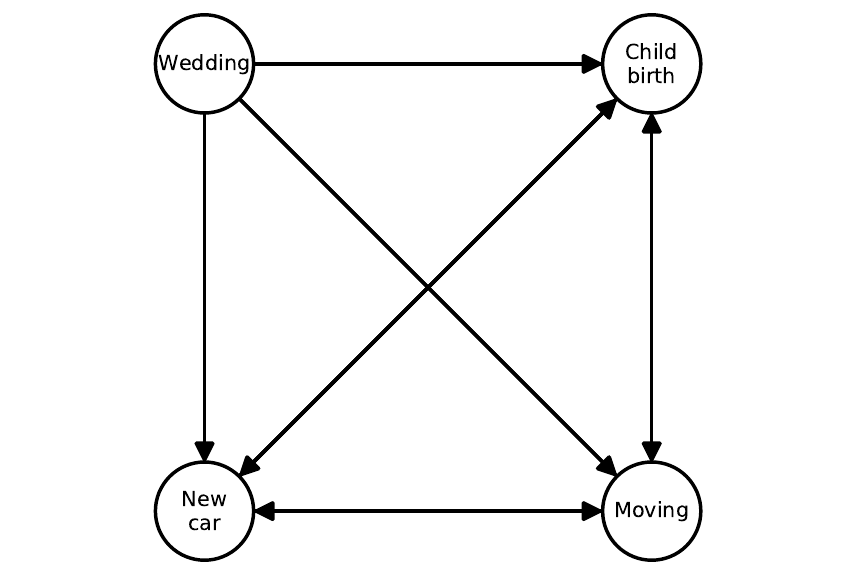}
    \caption{Final events graph containing four life events.}
    \label{fig:ges-reduced}
\end{figure}

The PC graph shows inconsistencies with respect to \emph{wedding} and \emph{child birth}, that we could not find in the dataset or in our personal intuition, particularly that \emph{child birth} precedes being \emph{married} (but not directly \emph{wedding}, which becomes an isolated event, only preceded by \emph{moving}). The child birth event almost certainly happens after other events (as can be seen in Figure A2), which was not uncovered by PC. It is apparent that the \emph{married} state variable confuses the algorithm. On the other hand, in the GES case, the relationships do not show contradiction with our intuition and the graph proposes more relationships. This is valuable from the perspective of our work, because it allows for more diversity in life-course trajectories. We therefore decided to keep working with the GES algorithm for the remainder of this paper. We recognize that this decision is certainly quite arbitrary, and it also reflects the well-known challenge of comparability of graphs in causal discovery \cite{singh2017comparative}. These algorithms themselves use different non-comparable score schemes \cite{causal_discovery_review}.

The last step in this process was to reduce the graph to only events variables; henceforth we will only use the graph in \ref{fig:ges-reduced}.

\subsection{Empirical Life Event Occurrence Distribution}

\begin{figure*}[ht]
    \centering
    \includegraphics[width=1.0\linewidth]{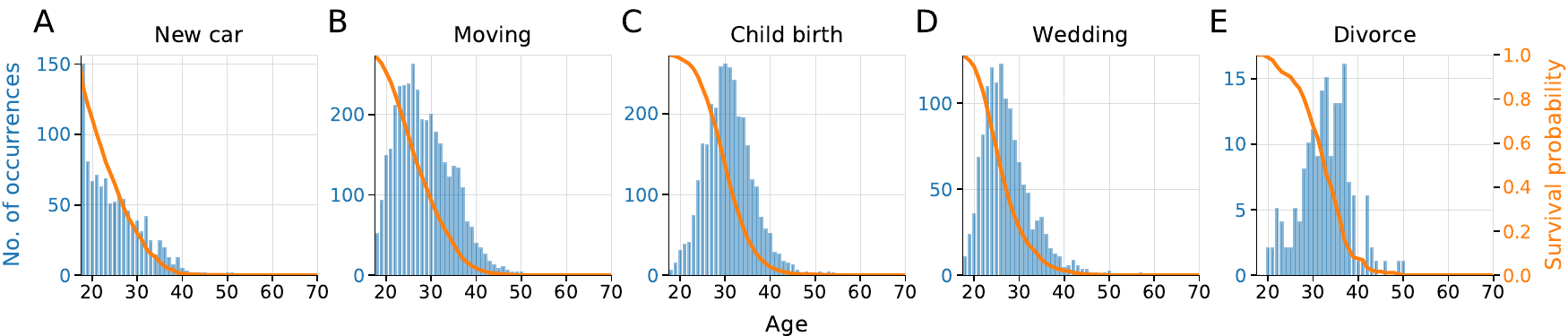}
    \caption{Empirical histograms and corresponding survival functions. The x-axis shows age, the left-hand side y-axis shows observed counts of specific event in the dataset (depicted in blue) and the right-hand side y-axis shows derived survival functions from the data (depicted in orange). Since all of our samples are fully observed (i.e. uncensored), all survival functions converge to a value of zero. (\textit{A}) New car event. (\textit{B}) Moving event. (\textit{C}) Child birth event. (\textit{D}) Wedding event. (\textit{E}) Divorce event.}
    \label{fig:hist-surv}
\end{figure*}

When a certain life event is most likely to happen in a person's life can be analysed from observed life-course data. Empirical histograms are a good way to understand occurrence distribution of events throughout the lifetime. Since in this case we work with completely observed data (only uncensored observations), the corresponding cumulative distribution functions (CDF), and thus the survival functions, can be directly derived. \ref{fig:hist-surv} shows histograms and survival functions for five life events: new car, moving, child birth, wedding and divorce.

The results show that buying a new car is likely to happen by the age of 40, with decreasing probability in time, with the age of 18 having the highest probability, which can be related to a legal minimum age for obtaining a driving license for unrestricted car use in Germany (\ref{fig:hist-surv}\textit{A}). Home relocation is most likely to happen in the later 20s, with the probability increasing up to the age of 25, and then decreasing up to the age of 50, after which people seem unlikely to move again (\ref{fig:hist-surv}\textit{B}). Having a child follows a normal distribution peaking around the age of 30, with a small chance of happening after the age of 45 (\ref{fig:hist-surv}\textit{C}). Getting married roughly follows a skewed distribution with a heavier tail on the increasing age side, and is most likely to happen in the 20s, specifically around mid 20s (\ref{fig:hist-surv}\textit{D}). Relating the events of getting married and having a child, having a child is shifted a couple of years after getting married, implying that in general having a child will likely happen soon after the wedding. Divorce is most likely to happen in the 30s, and generally before the age of 50, however the small sample size indicates an overall low probability of this happening (\ref{fig:hist-surv}\textit{E}).

\subsection{Describing General Life Event Transitions}

\begin{figure*}[h!]
    \centering
    \includegraphics[width=1.0\linewidth]{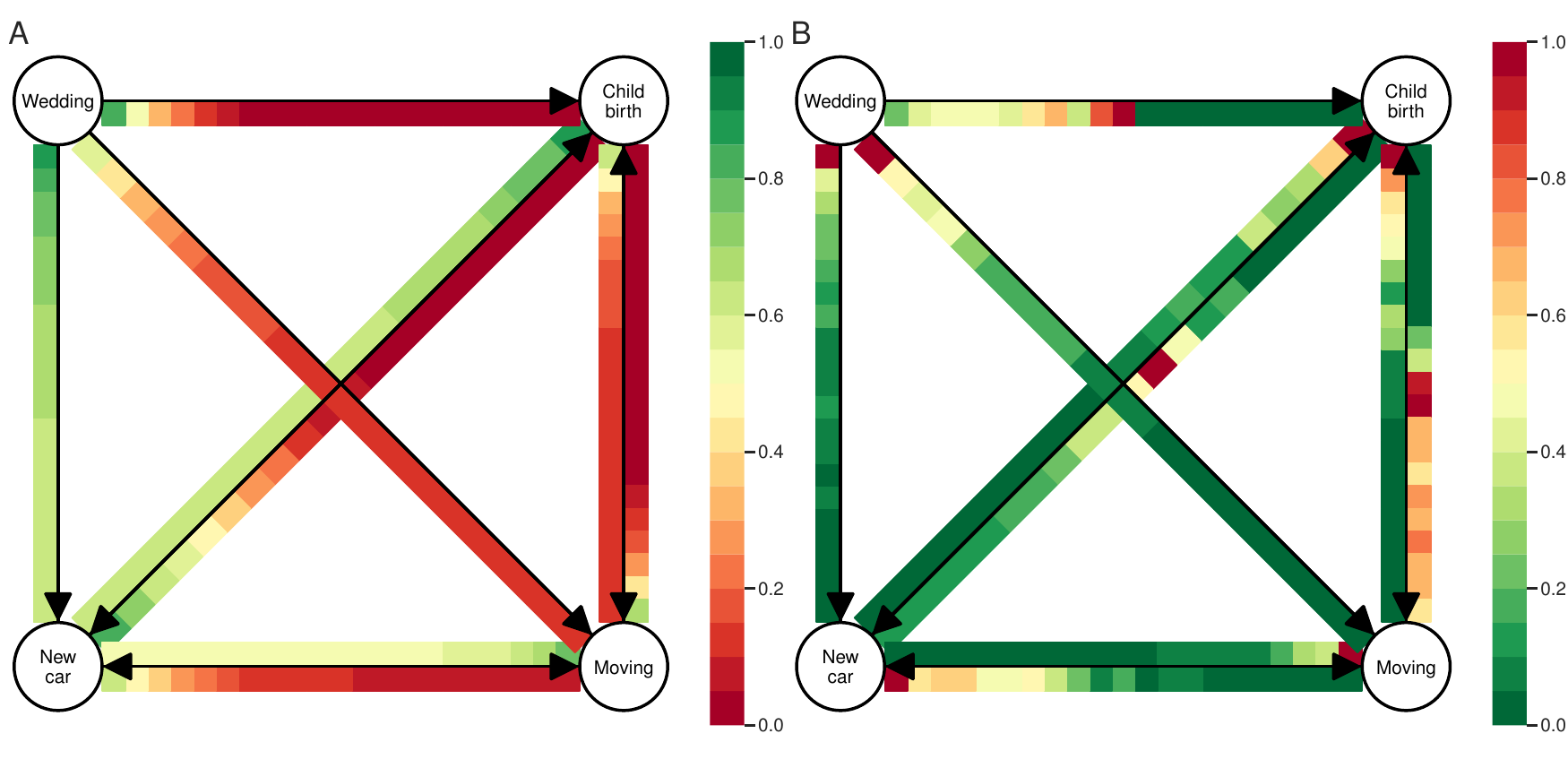}
    \caption{General life event transitions, represented in a life event graph. Nodes represent life events and directed edges represent life event pairs, with results shown on the right-hand side of the edges. Each edge is divided into 21 equal-length parts, representing a period of 20 years and the base year, with its color depicting the function value shown in the scale from the color bar on the side of each graph. (\textit{A}) Survival functions for each pair of life events. The function for each pair can be interpreted either individually or jointly compared among each other, as they all come from the same scale from \numrange{0}{1}. (\textit{B}) Normalized hazard functions derived from the cumulative hazard functions computed using the Nelson-Aalen method. These functions should be interpreted only individually, to compare hazard rates between intervals, and not regarded jointly, as they are normalized to a range from \numrange{0}{1} coming from different scales.}
    \label{fig:graphs-ges-surv}
\end{figure*}

Survival and cumulative hazard functions are estimated using Kaplan-Meier and Nelson-Aalen methods (\eqref{eq:km} and \eqref{eq:na}, respectively), and are shown in \ref{fig:graphs-ges-surv}. Kaplan-Meier estimates of the survival function are shown in \ref{fig:graphs-ges-surv}\textit{A}, where colors represent survival probabilities ranging from \numrange{0}{1}. The more the color at time $ t $ is on the green side, the higher the survival probability $ S(t) $ is, which indicates that there is a smaller probability that the event will happen before or at time $ t $. This can also be interpreted as showing that the event has a higher probability of surviving up to time $ t $. Normalized hazard functions are derived from the cumulative hazard function estimates and are shown in \ref{fig:graphs-ges-surv}\textit{B}, with a range from \numrange{0}{1}.

Looking at \ref{fig:graphs-ges-surv}\textit{A}, we can see that in the first 2-3 years after getting married, it is more likely that moving will happen next, rather than child birth or having a new car, which will then be superseded by having a child, as the survival probabilities decrease much faster after 3 years. Having a new car after getting married does not seem to be as important as child birth or moving. After having a child, moving is more likely than buying a new car. After moving, buying a car is unlikely compared to having a child, which is likely to happen within several years after moving. After buying a new car, there is a similar chance of moving as having a child.

\ref{fig:graphs-ges-surv}\textit{B} gives interesting insights regarding specific intervals when the event has a higher risk of happening on a timeline. For example, after getting married, the new car or moving event is very likely to happen within one year, whereas the child birth event has a higher risk of happening 10 or 11 years after getting married than in other periods. These estimates should be taken with caution: comparison of hazard rates among intervals gives individual difference between the rates. However, the hazard rates add up along the timeline, hence the cumulative risk of an event happening increases in time.

\subsection{The Effect of Covariates on Transition Probabilities}

\begin{figure*}[ht]
    \centering
    \includegraphics[width=1.0\linewidth]{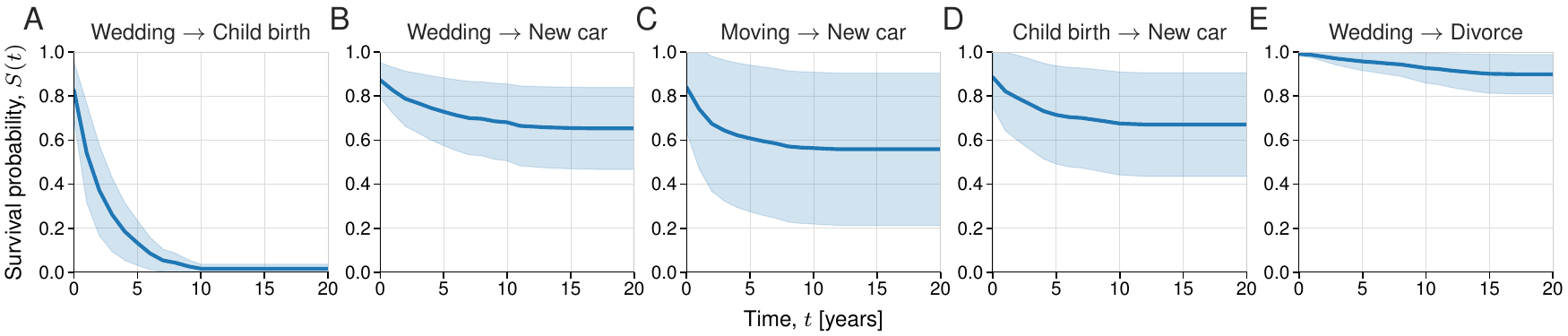}
    \caption{Average life event transitions behaviour for five pairs of life events. Predicted survival functions are computed for all dataset samples for each pair separately, which are then averaged, thus marginalizing over all the covariates to account for their empirical distributions. The blue line represents the survival function, and the light blue area around it is the confidence belt ($ \pm $ 1 standard deviation). The average survival functions have different shapes, with different confidence belts at different times. The narrower the confidence belt is, the less impact covariates have on the survival probability. Since not all of our samples are fully observed (some are censored as in some cases the event does not happen within the study period), survival functions do not end up at 0. (\textit{A}) Wedding to child birth transition. (\textit{B}) Wedding to new car transition. (\textit{C}) Moving to new car transition. (\textit{D}) Child birth to new car transition. (\textit{E}) Wedding to divorce transition.}
    \label{fig:surv-avg-functions}
\end{figure*}

The Cox proportional hazards models were trained and validated for each pair of life events from the life events graph. To understand the relationships in the data and prevent drawing false interpretations from the findings, detailed data preparation and description, as well as models training and validation, can be found in the previous section and in Appendix. Important to note is that the covariates obtained from time-dependent variables (e.g. age, number of children, number of cars, etc.) are measured at the time of the initial baseline event, which may help understand some of the less intuitive results shown later in the text (e.g. those without (or fewer) children at the time of wedding are less likely to get divorced, which can be explained by the fact that those are the ones who do not bring children into a partnership, and this results in the lower divorce probability, as shown in \ref{fig:surv-functions}\textit{W}). Another explanation comes from the small sample size of some covariates values (see Figure A4), e.g. people with more cars tend to buy cars faster than the ones with less or no cars, as shown later in \ref{fig:surv-functions}\textit{N}.

\subsubsection{Average life event transitions behaviour}

Average transition behaviour can give us an insightful overview of the survival behaviour within each pair and among the pairs, as shown in \ref{fig:surv-avg-functions}. Having a child is very likely to happen within 5 years from getting married, and almost certainly within ten years, with a probability close to 0 after 10 years. There is a small probability of having a child in the same year the wedding happens (\ref{fig:surv-avg-functions}\textit{A}). Buying a new car after getting married, as depicted in \ref{fig:surv-avg-functions}\textit{B}, has almost identical probability as buying a new car after having a child (\ref{fig:surv-avg-functions}\textit{D}), which will happen within the first 10 years from the initial event, and is unlikely after that. There is big variability in buying a new car after moving, demonstrated by the wide confidence interval in \ref{fig:surv-avg-functions}\textit{C}, with a similar switch point at around 10 years. The divorce event is unlikely to happen, with survival probability slightly decreasing in time over the period of within 15 years (\ref{fig:surv-avg-functions}\textit{E}).

It is important to note that general survival functions shown in \ref{fig:graphs-ges-surv}\textit{A} and average survival functions in \ref{fig:surv-avg-functions} differ in the way that the former are obtained directly from the data using univariate models, while the latter are obtained by averaging over all predicted survival functions using fitted multivariate CPH regression models, thus including the effect of covariates and the model error.

\subsubsection{Life event transitions behaviour}

\begin{figure*}[h!]
    \centering
    \includegraphics[width=1.0\linewidth]{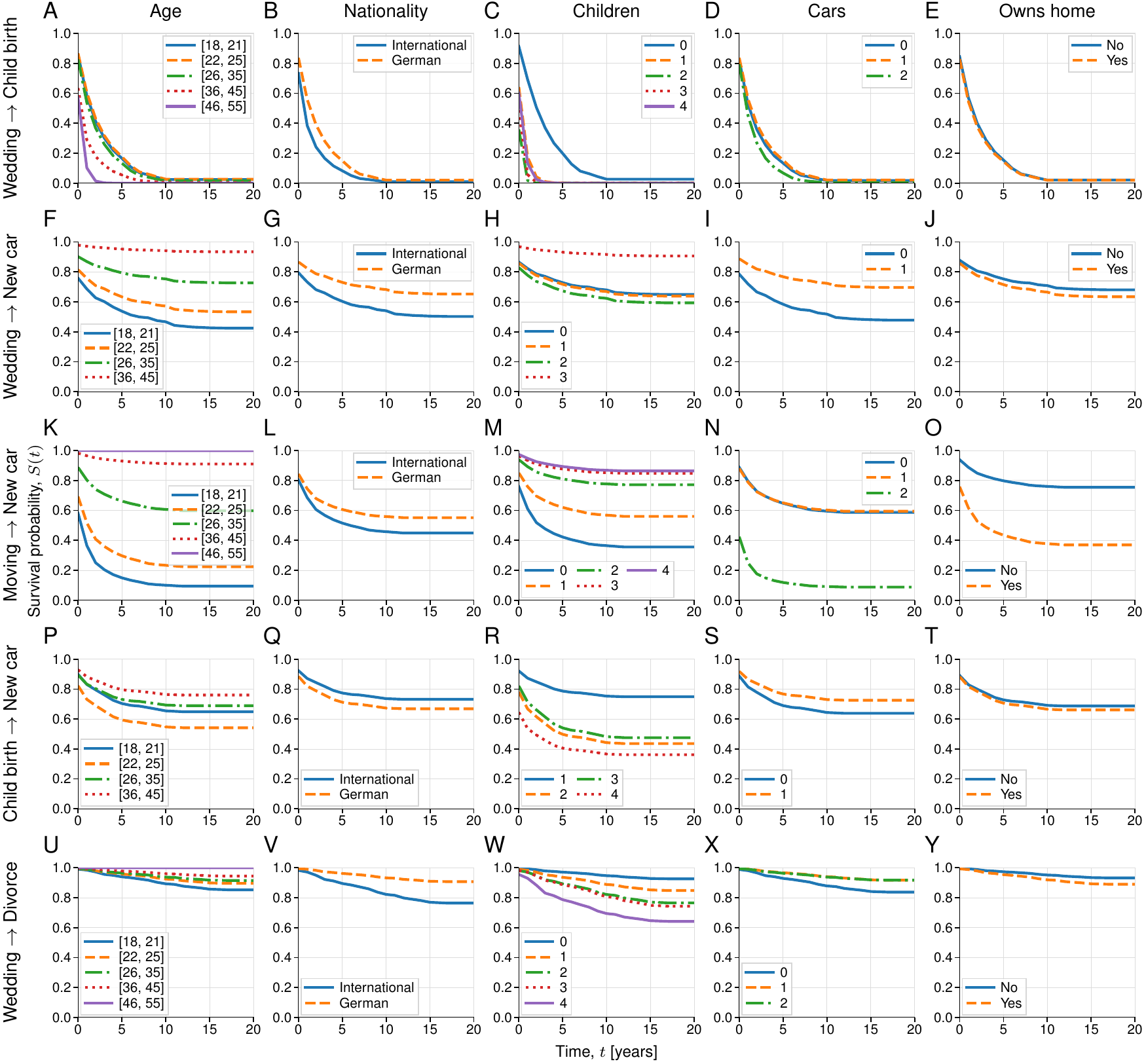}
    \caption{Life event transitions behaviour for five selected pairs of life events expressed through survival functions. Each row with five graphs represent behaviour for a specific life events pair, with five columns indicating which socio-demographic covariates the transition behaviour is conditioned on. For each graph, the x-axis represents time in years passed from the initial event, modelling the probability of another event happening. The covariates are age, nationality, number of children, number of cars and owning a home, respectively in order of plotting. \textit{(A-E)} Wedding to child birth transitions. \textit{(F-J)} Wedding to new car transitions. \textit{(K-O)} Moving to new car transitions. \textit{(P-T)} Child birth to new car transitions. \textit{(U-Y)} Wedding to divorce transitions.}
    \label{fig:surv-functions}
\end{figure*}

To examine the effect of socio-demographics on transition probabilities, we condition the prediction on one covariate for each transition pair (\ref{fig:surv-functions}).

\paragraph*{\textit{Wedding} to \textit{Child birth}}
The decision to have a child after getting married is mainly driven by two factors: age and the number of children (\ref{fig:surv-functions}\textit{A-E}). In terms of age, after wedding younger people tend to have children later, where the age groups of $ [36, 45] $ and $ [46, 55] $ tend to have children before the others, with a high chance of having a child in the same year as the wedding or likely within 3 years from it. Regardless of age, if a couple does not have a child within 10 years of getting married, they will probably not have a child at all. In terms of nationality, Germans tend to have children slightly later than other nationals. Interestingly, couples with no previous children tend to have children later than couples with children, who are likely to have a child within 1 to 2 years after getting married. Couples who already have children before the wedding could be couples who had children together before getting married or who had children with previous partners. The number of cars does not play a significant role in the decision to have a child. Similarly, whether a couple owns a home, which can be related to their economic situation, does not seem to be relevant for having a child.

\paragraph*{\textit{Wedding} to \textit{New car}}
The decision to have a new car after getting married is driven by several factors, with the most influential being age and the current number of cars (\ref{fig:surv-functions}\textit{F-J}). The youngest age group $ [18, 21] $ tend to have a new car sooner than the others, with the probability gradually decreasing with increases in age, where people in the age group $ [36, 45] $ are very unlikely to have a new car after getting married. Germans tend to have a new car later than other nationals. Those who have no children or one child exhibit similar behaviour, while having two children speeds up the decision to buy a car, whereas couples with three children postpone it. Intuitively, couples with no car tend to buy a car much sooner than couples who already possess a car. Couples who own a home tend to buy a new car slightly sooner then couples who do not own a home, which may be related to a better economic situation.

\paragraph*{\textit{Moving} to \textit{New car}}
The decision to have a new car after moving is driven by all factors except nationality (\ref{fig:surv-functions}\textit{K-O}). The difference in behaviour between age groups is very distinguishable - the younger the person the more likely he/she is to have a new car after moving, with the youngest group $ [18, 21] $ tending to have a car within 2 years from moving. Germans tend to buy a new car later than other nationals. The number of children plays an important role, where people with fewer children tend to by a car sooner, whereas people with 3 or more children are less likely to buy a new car after moving. For people who had no children at the time of moving, they are more likely (compared to others) to have a new (first) child after moving, which in turn may explain the faster car acquisition. Surprisingly, people living in households with two cars or more are very likely to buy or replace a car soon after moving, most of them within two years, including the year of moving. The magnitude of this is very strong, which may be an artifact of a small sample group, or the different possibilities in temporal coding. People who own a home are significantly more likely to get a car before people who do not own a home.

\paragraph*{\textit{Child birth} to \textit{New car}}
The decision of having a new car after having a child is driven by several factors, of which the most influential are age and the number of cars (\ref{fig:surv-functions}\textit{P-T}). With the exception of the youngest age group $ [18, 21] $, the increase in age postpones buying a car after having a child. Germans are likely to have a new car slightly sooner than other nationals. People with one child will probably buy a car much later than people with more than one child. In line with intuition, after having a child people without a car tend to buy a car sooner than people with a car. Similarly, people who own a home are more likely to buy a car than people not owning a home are.

\paragraph*{\textit{Wedding} to \textit{Divorce}}
The overall probability of getting divorced is low, but some variability is still present (\ref{fig:surv-functions}\textit{U-Y}). In general, based on observed divorces, if a divorce does not happen within 15 years from the wedding, it will probably not happen. The youngest age group tend to get divorced sooner than others. Germans tend to get divorced later than other nationals. People with fewer children have lower divorce probability than people with more children, with a sensitive group being people with four children. An intuitive interpretation for this could be that those without (or fewer) children at the time of getting married are those that do not bring children into a partnership, and this results in the lower divorce probability. The number of cars can also affect divorce probability, where people with fewer cars are more likely to get divorced. Homeowners are more likely to get a divorce than non-homeowners.

\subsubsection*{The effect of socio-demographics on transitions}

Regarding the general effect of socio-demographics on transitions, younger age groups generally prefer to have a new car and divorce sooner, while transitions to having a child is faster as age increases, with the oldest group tending to postpone the decision of buying a car (\ref{fig:surv-functions}\textit{A,F,K,P,U}). Regarding nationality, other nationals make faster transitions than Germans, except when it comes to buying a car after having a child (\ref{fig:surv-functions}\textit{B,G,L,Q,V}). The number of children seems to play a role in all transitions (\ref{fig:surv-functions}\textit{C,H,M,R,W}). The number of cars, as expected, tends to have an effect in the decision to buy a new car (\ref{fig:surv-functions}\textit{D,I,N,S,X}). Homeowners generally have faster transitions than people not owning a home (\ref{fig:surv-functions}\textit{E,J,O,T,Y}).

\section{Conclusion}
\label{sec:conclusion}

Numerous studies in transport research and sociology seek to model the structure of the life course and the inter-connection between life-course events (birth of a child, home relocation, car change, etc.). The use of traditional statistical models does not generally allow an analyst to infer causality without overrelying on expert knowledge. Indeed, statistical models of choice, which are predominantly used in the field, provide correlation measures rather than causation measures. By introducing a bi-level model where the upper level is a causal discovery algorithm and the lower level is composed of a series of survival analyses modelling the transition rate between life course events, we introduce a more rigorous data driven assessment of causality in life-course analysis. The results of this research can be used for high-level activity modelling for mobility applications, to better understand how life events and socio-demographic characteristics affects and constraints people's mobility and travel choices.

\setlength{\parindent}{0cm}

\section*{Acknowledgements}
\label{sec:acknowledgements}

The research leading to these results has received funding from the European Union's Horizon 2020 research and innovation program under the Marie Sklodowska-Curie grant agreement no. 754462. The Leeds authors acknowledge the financial support by the European Research Council through the consolidator grant 615596-DECISIONS.

\clearpage


\bibliographystyle{ieeetr}
\bibliography{cite}

\clearpage
\setlength{\parindent}{1cm}

\renewcommand*{\thefigure}{A\arabic{figure}}
\renewcommand*{\thetable}{A\arabic{table}}
\setcounter{figure}{0}

\section*{Appendix}
\label{sec:appendix}

\subsection*{Data collection}
\label{subsec:app_data_collection}

A screenshot of the life-course calendar used in this study is shown in Fig. A1.

\begin{figure}[hb]
    \centering
    \includegraphics[width=0.75\textwidth]{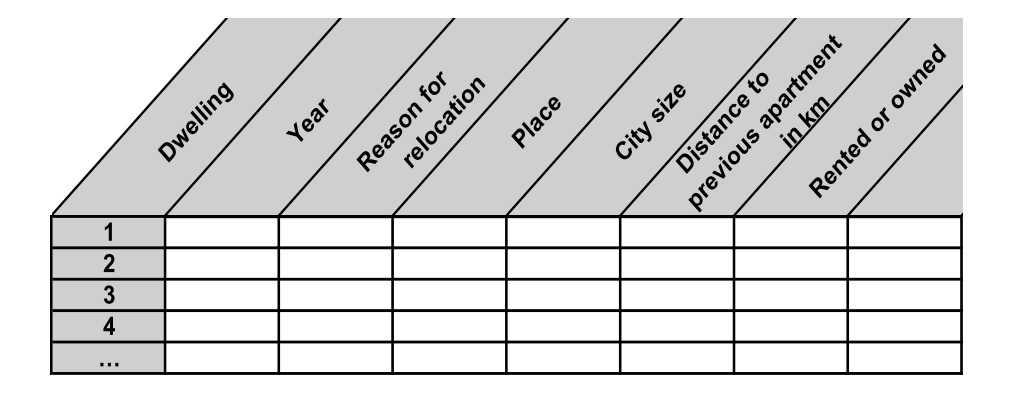}
    \caption{A screenshot of the life-course calendar used in this study (translated from German into English.}
    \label{fig:att-lit-lcc}
\end{figure}

\subsection*{Causal discovery}
\label{subsec:app_causal_discovery}

In order to prepare the data for the algorithms, we created an observation for each event pair occurrence, as illustrated in Table A1. Observation 2 can be read this way: ``at some moment in time, a married individual with 2 children had a child birth event, which was followed by a moving event later in life''. Due to the nature of the dataset, an ``event occurrence'' always matches a year, so one may have multiple events preceding multiple events, as happens in observation 4.

\begin{sidewaystable}
    \centering
    \caption{An example of four \emph{event pair} observations.}
    \label{tab:app-event-obs}
    \begin{tabular}{ *{7}{c} *{4}{p{1.3cm}}}
        ID & New car & Moving & Child birth & Wedding & Married & Children & New car next event & Moving next event & Child birth next event & Wedding next event \\
        \midrule
        1 & 0 & 1 & 0 & 0 & 1 & 2 & 1 & 0 & 0 & 0 \\
        2 & 0 & 0 & 1 & 0 & 1 & 2 & 0 & 1 & 0 & 0 \\
        3 & 1 & 0 & 1 & 0 & 1 & 1 & 0 & 0 & 1 & 0 \\
        4 & 1 & 1 & 0 & 1 & 1 & 0 & 1 & 0 & 1 & 0 \\
        \bottomrule
    \end{tabular}
\end{sidewaystable}

Regarding the choice of the causal graph, Fig. A2 shows the survival probability marginalized over all other events (new car, moving, wedding) leading to the child birth event for the GES graph. There is a clear indication that the child birth happens almost certainly after each one of these events. These relations are not present in the PC graph, where only the new car event precedes the child birth event. 
\begin{figure}
    \centering
    \includegraphics[width=0.38\linewidth]{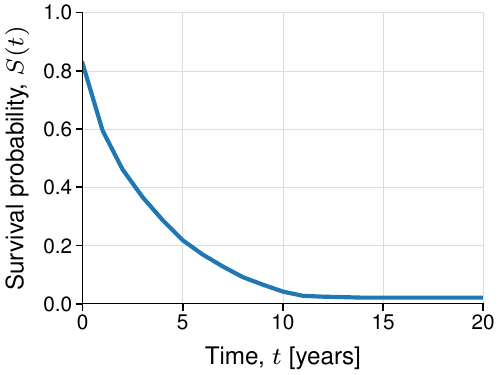}
    \caption{Average survival curve from all other events (new car, moving, wedding) to the child birth event.}
    \label{fig:surv-avg-birth-marg}
\end{figure}

\subsection*{Survival analysis}
\label{subsec:app_survival_analysis}

Fig. A3 shows a general overview of observations in survival analysis.

\begin{figure}
    \centering
    \includegraphics[width=0.65\linewidth]{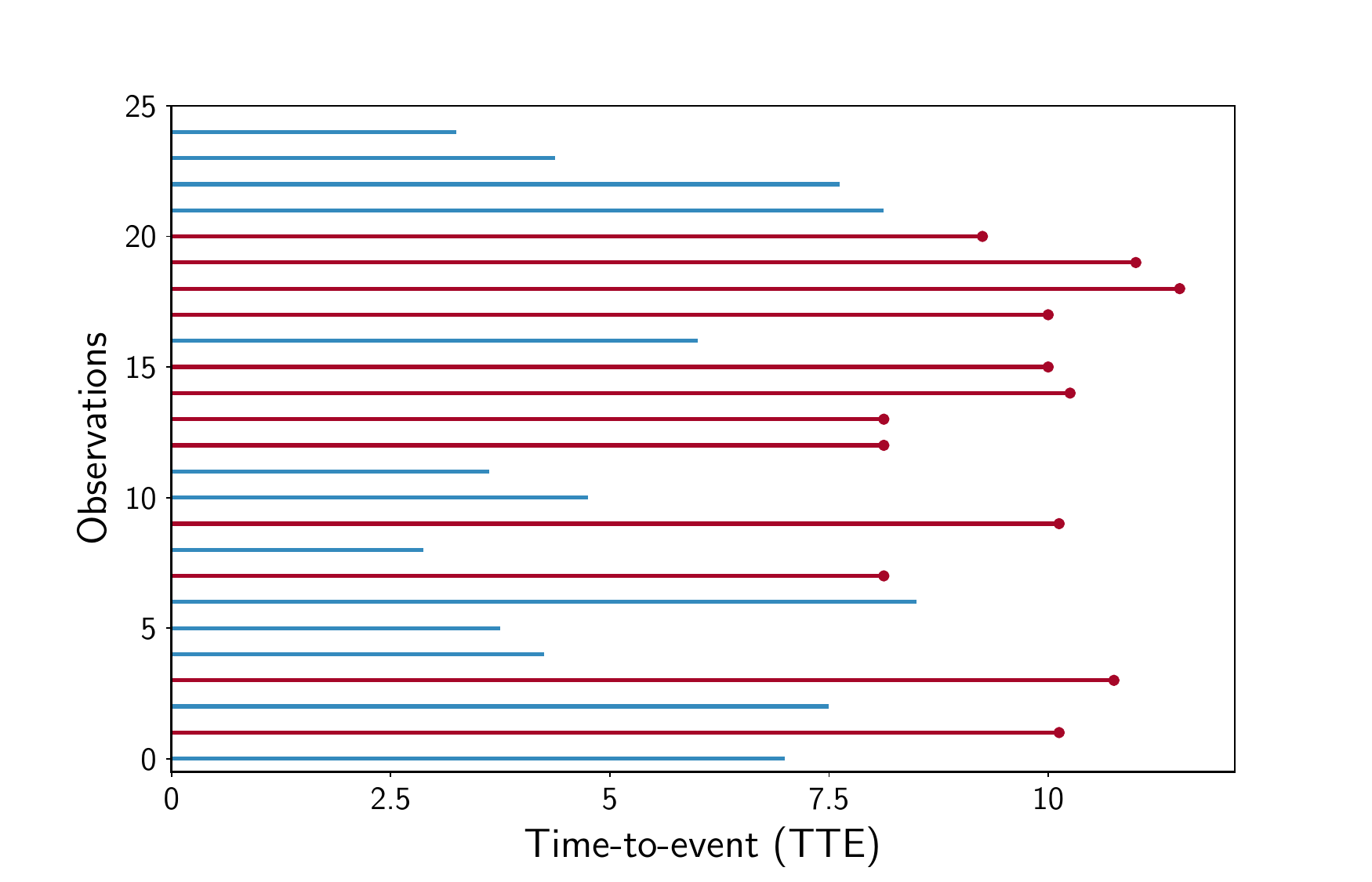}
    \caption{General example of survival analysis data. Uncensored observations (red) and censored observations (blue).}
    \label{fig:app-surv-data}
\end{figure}

\subsection*{Data elaboration}
\label{subsec:app_data_elaboration}

An example of elaborated time-dependent data for one person from the dataset used for survival analysis is shown in Table A2.

\begin{sidewaystable}
    \centering
    \caption{An example of time-dependent data for one person from the dataset.}
    \label{tab:app-example-person}
    \begin{tabular}{ l *{20}{c}}
        \multirow{2}{*}{Attribute} & \multicolumn{20}{c}{Year} \\ \cline{2-21}
         & 1 & 2 & 3 & 4 & 5 & 6 & 7 & 8 & 9 & 10 & 11 & 12 & 13 & 14 & 15 & 16 & 17 & 18 & 19 & 20 \\
        \midrule
        Age & 18 & 19 & 20 & 21 & 22 & 23 & 24 & 25 & 26 & 27 & 28 & 29 & 30 & 31 & 32 & 33 & 34 & 35 & 36 & 37 \\
        Owns home & 1 & 1 & 1 & 1 & 1 & 1 & 1 & 1 & 1 & 1 & 1 & 1 & 1 & 1 & 1 & 1 & 1 & 1 & 1 & 1 \\
        Cars & 0 & 0 & 0 & 0 & 0 & 0 & 0 & 0 & 1 & 1 & 1 & 1 & 1 & 1 & 1 & 1 & 1 & 1 & 1 & 1 \\
        Distance to work & 0 & 0 & 8 & 7 & 7 & 5 & 5 & 5 & 5 & 5 & 5 & 7 & 7 & 7 & 7 & 7 & 7 & 7 & 7 & 7 \\
        Rides car & 0 & 0 & 0 & 0 & 0 & 0 & 0 & 0 & 0 & 0 & 0 & 0 & 0 & 0 & 0 & 0 & 0 & 0 & 0 & 0 \\
        Children & 0 & 0 & 0 & 0 & 0 & 0 & 0 & 1 & 1 & 1 & 1 & 1 & 1 & 1 & 1 & 1 & 1 & 1 & 1 & 1 \\
        Married & 0 & 0 & 0 & 0 & 0 & 0 & 1 & 1 & 1 & 1 & 1 & 1 & 1 & 1 & 1 & 1 & 1 & 1 & 1 & 1 \\
        New car & 0 & 0 & 0 & 0 & 0 & 0 & 0 & 0 & 1 & 0 & 0 & 0 & 0 & 0 & 0 & 0 & 0 & 0 & 0 & 0 \\
        Moving & 0 & 0 & 0 & 0 & 0 & 0 & 0 & 0 & 0 & 0 & 0 & 0 & 0 & 0 & 0 & 0 & 0 & 0 & 0 & 0 \\
        Child birth & 0 & 0 & 0 & 0 & 0 & 0 & 0 & 1 & 0 & 0 & 0 & 0 & 0 & 0 & 0 & 0 & 0 & 0 & 0 & 0 \\
        Wedding & 0 & 0 & 0 & 0 & 0 & 0 & 1 & 0 & 0 & 0 & 0 & 0 & 0 & 0 & 0 & 0 & 0 & 0 & 0 & 0 \\
        Divorce & 0 & 0 & 0 & 0 & 0 & 0 & 0 & 0 & 0 & 0 & 0 & 0 & 0 & 0 & 0 & 0 & 0 & 0 & 0 & 0 \\
        \bottomrule
    \end{tabular}
\end{sidewaystable}

\subsection*{Covariates description}
\label{subsec:app_covariates_description}

The covariates are divided into five groups: general attributes, independent state attributes, dependent state attributes, transition attributes and indicator transition attributes. Their description, type and feasible values are shown in Table A3.

General attributes are \textit{Gender}, \textit{Nationality}, \textit{No. of relocations} and \textit{City size}. They represent general characteristics of individuals. Their value neither depends on the year (they are time-constant), nor on the life events pair. \textit{Gender} and \textit{Nationality} are categorical attributes, \textit{No. of relocations} and \textit{City size} are discrete.

Independent state attributes are \textit{Age}, \textit{Owns home}, \textit{Distance to work} and \textit{Rides car}. These are time-dependent attributes. Given a life events pair, an independent state attribute takes a value it has in the year the cause event occurs. \textit{Age}, \textit{Owns home} and \textit{Rides car} are categorical, whereas \textit{Distance to work} is discrete (discretised to 1 km).

Dependent state attributes are \textit{No. of cars}, \textit{No. of children} and \textit{Married}. They are affected by the transition attributes. Their value is computed in the same way as for the independent state attributes. They are all categorical.

Transition attributes are \textit{New car}, \textit{Moving}, \textit{Child birth}, \textit{Wedding} and \textit{Divorce}. These are binary variables, having values of mostly 0, and 1 when the event happens. They increment a dependent state attribute (in case of \textit{No. of cars} and \textit{No. of children}), or alternate the value of \textit{Married}. These attributes are treated in a different way. To be used in estimations, they are expanded in a way how many years they happened before the cause event. Therefore, they take values in the range $ [0, -19] $, since the dataset for each person covers a period of 20 years. They are all discrete.

Indicator transition attributes are dummy variable used to account for whether the transition attribute happened or not before the cause effect. They are all categorical.

\begin{sidewaystable}
    \centering
    \caption{Covariates description}
    \label{tab:app-covariates-description}
    \begin{tabular}{ *{5}{l} }
        Group & Name & Description & Type & Domain \\
        \midrule
        
        \multirow{4}{*}{General attributes} & Gender & gender, is the person male or female & categorical & \{0, 1\} \\
         & Nationality & nationality, is the person German or international & categorical & \{0, 1\} \\
         & No. of relocations & no. of times the person relocated with parents & discrete & [0, 9] \\
         & City cize & the size of the city the person lives in & discrete & [0, 7] \\
        \midrule
        
        \multirow{4}{*}{Independent state attributes} & Age & age, a range of 20 years for every person & categorical & 5 age groups \\
         & Owns home & does the person own a home or not & categorical & \{0, 1\} \\
         & Distance to work & travelling distance to work [km] & discrete & [0, 100] \\
         & Rides car & if the person uses the car to go to work & categorical & \{0, 1\} \\
        \midrule
        
        \multirow{3}{*}{Dependent state attributes} & No. of cars & the number of cars the person has & categorical & [0, 2] \\
         & No. of children & the number of children the person has & categorical & [0, 3] \\
         & Married & marital status, is the person married or not & categorical & \{0, 1\} \\
        \midrule
        
        \multirow{5}{*}{Transition attributes} & New car & the event of buying a new car & discrete & $ [0, -19] $ \\
         & Moving & the event of moving & discrete & $ [0, -19] $ \\
         & Child birth & the event of having a child & discrete & $ [0, -19] $ \\
         & Wedding & the event of having a wedding (getting married) & discrete & $ [0, -19] $ \\
         & Divorce & the event of having a divorce & discrete & $ [0, -19] $ \\
        \midrule
        
        \multirow{5}{*}{Indicator transition attributes} & Indicator - New car & indicator for the event of buying a new car & categorical & \{0, 1\} \\
         & Indicator - Moving & indicator for the event of moving & categorical & \{0, 1\} \\
         & Indicator - Child birth & indicator for the event of having a child & categorical & \{0, 1\} \\
         & Indicator - Wedding & indicator for the event of having a wedding (getting married) & categorical & \{0, 1\} \\
         & Indicator - Divorce & indicator for the event of having a divorce & categorical & \{0, 1\} \\
        \bottomrule
    \end{tabular}
\end{sidewaystable}

Distribution of covariates values is shown in Fig. A4.

\begin{figure}
    \centering
    \includegraphics[width=1.\linewidth]{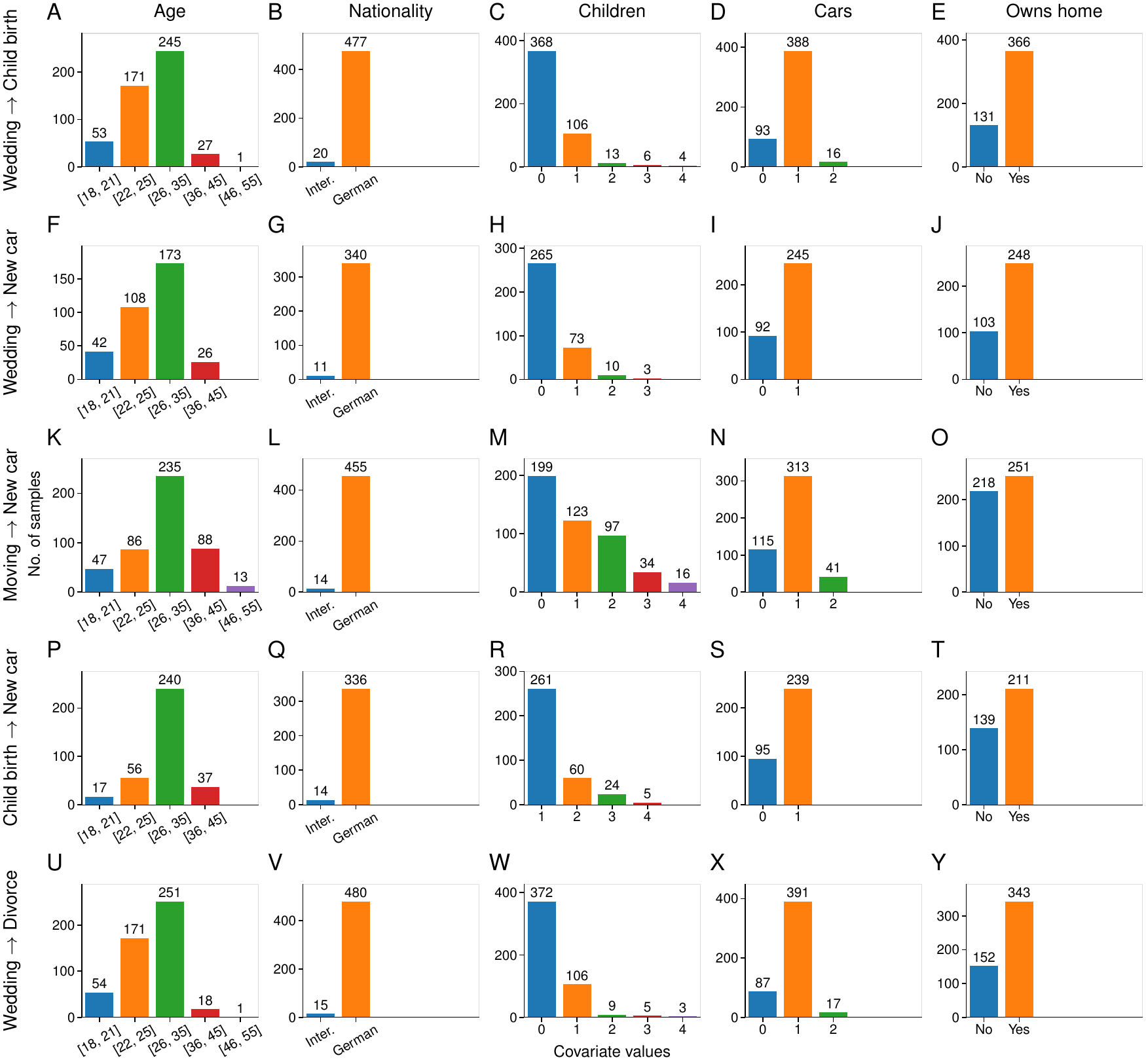}
    \caption{Distribution of covariates values.}
    \label{fig:app-surv-data-hist}
\end{figure}

\subsection*{Model fitting}
\label{subsec:app_model_fitting}

Model fitting results are shown in Table A4.

\begin{table}
    \centering
    \caption{Fitted CPH models}
    \begin{tabular}{ llrrr }
        Cause & Effect & Sample size & C-index - train & C-index - test \\
        \midrule
        New car & Moving & 819 & 0.687 & 0.605 \\
        New car & Child birth & 736 & 0.801 & 0.757 \\
        Moving & New car & 1172 & 0.863 & 0.864 \\
        Moving & Child birth & 1704 & 0.831 & 0.824 \\
        Child birth & Moving & 1481 & 0.670 & 0.679 \\
        Child birth & New car & 873 & 0.784 & 0.809 \\
        Wedding & Moving & 1175 & 0.591 & 0.604 \\
        Wedding & Child birth & 1241 & 0.733 & 0.748 \\
        Wedding & Divorce & 1237 & 0.719 & 0.682 \\
        Wedding & New car & 877 & 0.608 & 0.652 \\
        \bottomrule
    \end{tabular}
    \label{tab:app-fitted-coxph}
\end{table}

\end{document}